\documentclass[preprintnumbers,prb,superscriptaddress,twocolumn,floatfix,reprint,10pt]{revtex4-1}
\usepackage[T1]{fontenc}
\usepackage[latin9]{inputenc}
\usepackage{textcomp}
\usepackage{amstext}
\usepackage[Symbol]{upgreek}
\usepackage{graphicx}
\usepackage{amsmath}
\usepackage{booktabs}
\usepackage{hyperref}
\usepackage{dcolumn}
\usepackage{enumerate}
\usepackage{latexsym,bm}
\makeatletter

\begin{document}

\title{Hybrid functional with semi-empirical van der Waals study of native defects in hexagonal BN}

\author{V. Wang}
\thanks{Corresponding author at: Department of Applied Physics, Xi'an University of Technology, No.58, Yanxiang Road, Xi'an 710054, China, Tel./Fax: +86-29-8206-6357/6359 \\E-mail address: wangvei@icloud.com (V. Wang).}
\affiliation{Department of Applied Physics, Xi'an University of Technology, Xi'an 710054, China}

\author{R.-J. Liu}
\affiliation{Department of Applied Physics, Xi'an University of Technology, Xi'an 710054, China}

\author{H.-P. He}
\affiliation{Department of Geological Engineering, Lanzhou Resources \& Environment Voc-Tech College,
Lanzhou 730021, China}

\author{C.-M. Yang}
\affiliation{Department of Applied Physics, Xi'an University of Technology, Xi'an 710054, China}


\author{L. Ma}
\affiliation{Department of Applied Physics, Xi'an University of Technology, Xi'an 710054, China}

\date{\today}

\begin{abstract}
\centerline{\bf{ABSTRACT}}
The formation energies and transition energy levels of native defects in hexagonal BN have been studied 
by first-principles calculations based on hybrid density functional theory (DFT) together with an empirical dispersion correction of Grimme's DFT-D2 method. Our calculated results predict that the interstitial B is the most stable defect under N-rich and \emph{p}-type conditions. While the B vacancy and interstitial N become the dominate defects when the electron chemical potential is near the conduction band maximum of host. Nevertheless, these compensating defects will be inactive due to their ultra deep ionization levels under both both \emph{p}- and \emph{n}-type conditions.
\end{abstract}

\keywords{D. First-principles calculation; C. Native defect; A. Hexagonal boron nitride}


\maketitle 

\section{introduction}
Boron nitride exists in various phases, including cubic (\emph{c}-BN), wurtzite (\emph{w}-BN), hexagonal (\emph{h}-BN) structures. Similar to graphite, \emph{h}-BN consists of stacked BN layers with equal numbers of boron and nitrogen atoms. Within each \emph{h}-BN layer, alternating boron and nitrogen atoms form a honeycomb sheet by \emph{sp}$^2$-hybridization; while the interlayer attractive force is mediated by weak van der Waals interactions. 
In contrast to the semimetallic behavior of graphite, \emph{h}-BN is a wide band-gap semiconducting material 
due to its partially ionic character of B-N bond.
Because it has high temperature stability, low dielectric constant, high mechanical strength, large thermal conductivity, high hardness, and high corrosion resistance. 
It is a promising material for the realization of compact ultraviolet laser devices, high-temperature and high-pressure devices. \cite{Watanabe2004,Kubota2007,Watanabe2009,Eichler2008} Furthermore, it can be well integrated with graphene to design new novel devices  
as they have small mismatch (1.6\%) and the same hexagonal structure. \cite{Dean2010, Britnell2012}

\emph{h}-BN is commonly synthesized through mechanical,\cite{Novoselov2005} liquid-phase exfoliation,\cite{Coleman2011} or chemical vapor deposition (CVD). \cite{Oshima1997,Corso2004,Shi2010,Song2010}
During the process of \emph{h}-BN growth, native defects are thus readily formed in an uncontrolled
way, and they can be unintentionally generated to modify the conductivity of \emph{h}-BN.  Recently, Shi \emph{et al.} synthesized \emph{h}-BN films with the thickness up to 20 $\mu$m on the Ni surface by CVD. They pointed that the B/N ratio reaches to be 1:1.12 detected by X-ray photoelectron spectroscopy.\cite{Shi2010}
Previous theoretical studies have investigated the native defects in \emph{h}-BN  based on traditional density functional theory (DFT) within generalized gradient approximation (GGA) or local density approximation (LDA).\cite{Orellana2001,Hou2009} The native defect properties of \emph{h}-BN bilayer and monolayer have also been studied in our earlier DFT-GGA calculations.\cite{Wang2012}
However, the severe underestimation of traditional DFT on the band gap of semiconductors, especially in the wide-gap semiconductors, results in large uncertainties in the position of defect levels and the stability of defects. \cite{Mosuang2002,Orellana2001,Zobelli2007,Azevedo2007,Okada2009,Yin2010,Alem2011} 
The recent development of hybrid density functional theory,\cite{Heyd2003} which mix a fraction of Hartree-Fock (HF) exchange with the LDA or GGA
exchange and correlation potentials, can correct the band gap \cite{Marsman2008} and provide more reliable description of transition levels and formation energies of defect in semiconductors.\cite{Lyons2009,VandeWalle2011,Alkauskas2011,Lyons2012}

In this paper, we have systematically investigated the formation energies and transition energies of native defects in \emph{h}-BN using hybrid density functional theory. We found that the calculated transition energies of all native defects are ultra deep and thus neither of them cannot contribute or compensate specific conductivity by extrinsic doping.    
The remainder of this paper is organized as follows.
In Sec. II, the details of the computation are described.
Sec. III presents our calculated formation energies and single particle energy levels of native defect in bulk hexagonal BN.
Finally, a short summary is given in Sec. IV.

\section{methods}
Our total energy and electronic structure calculations were based on spin-polarized Kohn-Sham theory with the hybrid functional proposed by Heyd, Scuseria, and Ernzerhof (HSE) \cite{Heyd2003}, employing the projector augmented wave potentials\cite{PAW} as
implemented in the VASP code.\cite{Kresse1994,Kresse1999} 
In the HSE approach, a screening parameter of 0.2 {\AA}$^{-1}$ was used as suggested for the HSE06 functional.\cite{Krukau2006} We found that a proportion
of $\alpha$=33\% HF exchange with 67\% GGA of Perdew, Burke and Ernzerhof (PBE) \cite{Perdew1996} exchange produces an accurate value of the band gap for \emph{h}-BN. The week van der Waals interaction between layers plays a key role in determining the interlayer distance for layered materials. We incorporated the van der Waals interactions through employing an empirical correction scheme of Grimme's DFT-D2 method, which has been proven to be successful in describing the geometries of layered materials. \cite{Grimme2006,Bucko2010} 
\emph{h}-BN has five possible structures. More details about the effects of stacking behavior can be found in Refs. \onlinecite{Liu2003, Marom2010}. The most stable AB-like stacking sequence with each boron atom on top of a nitrogen atom was adopted in our current study. 
The DFT-D2 (global scaling factor \emph{s}$_6$=0.40) plus HSE06 ($\alpha$=0.33) scheme gives the calculated optimized lattice constants  \emph{a}=2.50 {\AA} and \emph{c}=6.42 {\AA}, in good agreement with the experimental values of \emph{a}=2.50 and \emph{c}=6.66 {\AA} respectively.\cite{Pease1952}

The defect systems were modeled by adding (removing) an atom to (from) in a 4$\times$4$\times$2 supercell consisting of 128 atoms.
The wave functions were expanded by plane waves up to a cutoff energy of 300 eV. The integrations over the Brillouin zone were performed using $\Gamma$-centered 2$\times$2$\times$2 k-point mesh generated by Monkhorst-Pack scheme. \cite{Monkhorst1976} The internal coordinates in the defect supercells were relaxed to reduce the residual force to less than 0.02 eV$\cdot${\AA}$^{\text{-1}}$.
In the charged-defect calculations, a uniform background charge was added to keep the global charge neutrality of supercell.
The formation energy of a charged defect was defined as \cite{Zhang1991}

\begin{equation}\label{eq1}
\begin{split}
\Delta E^f_D(\alpha,q)=E_{tot}(\alpha,q)-E_{tot}(0)-\sum n_{\alpha}\mu_{\alpha}  \\ 
+q(\mu_{e}+\epsilon_{v}+\Delta V_{align}[q]),
\end{split}
\end{equation}

where $E_{tot}(\alpha,q)$ and $E_{tot}(0)$ are the total energies of the supercells with and without defect $\alpha$. \emph{n}$_\alpha$ is the number of species $\alpha$ ($\alpha$=B, N) needed to create the defect $\alpha$. $\mu_{\alpha}$ is the corresponding atomic chemical potential, and \emph{q} is the charge state of defect. $\mu_{e}$ is electron chemical potential in reference to the valence band maximum (VBM, denoted as $\epsilon_{v}$) of bulk \emph{h}-BN. Therefore, the electron chemical potential varies between zero and the band-gap \emph{E}$_g$ of host. 
The potential offset $\Delta$\emph{V}$_{align}[q]$ is determined by the difference of the atomic-sphere-averaged electrostatic potentials around host atoms father away the defect $\alpha$, relative to the atomic-sphere-averaged electrostatic potentials around host atoms in the defect-free supercell.\cite{Laks1992,Walle2004}

The chemical potential $\mu_{\alpha}$ depends on the experimental growth conditions. The chemical potentials of boron and nitrogen are subject
to their lower bounds satisfied by the constraint $\mu_{\text{BN}}(bulk)$=$\mu_{\text{B}}$+$\mu_{\text{N}}$, where $\mu_{\text{BN}}(bulk)$= is the total energy of per formula unit of \emph{h}-BN. 
The calculated formation energy of \emph{h}-BN is -2.95 eV, slightly unstable 0.35 eV than the value obtained by the previous GGA-based calculation.\cite{Orellana2001}
The $\mu_{\text{N}}$ is subject to an upper bound given by the energy of one N atom (-10.34 eV) in a N$_2$ molecule, which corresponds to extreme N-rich/B-poor growth condition. Similarly, the upper bound of the $\mu_{\text{B}}$ is limited by the energy of one B atom (-7.65 eV) in boron bulk phase, corresponding to extreme B-rich/N-poor growth condition. 
   
The transition-energy level $\epsilon_{\alpha}$(\emph{q}/$\emph{q}^{\prime}$) of defect is the Fermi energy in Eq. (\ref{eq1}) at which the formation energy $\Delta E^f_D(\alpha,\emph{q})$ of defect $\alpha$ with charge \emph{q} is equal to that of another charge $\emph{q}^{\prime}$ of the same defect. The  $\epsilon_{\alpha}$(\emph{q}/$\emph{q}^{\prime}$) can be calculated as

\begin{equation}\label{eq2}
\epsilon_{\alpha}(q/q^{\prime})=[\Delta E^f_D(\alpha,q)-\Delta E^f_D(\alpha,q^{\prime})]/(q^{\prime}-q).
\end{equation}

\section{Results and discussion}
\subsection{Pristine bulk \emph{h}-BN}
\begin{figure}[htbp]
\centering
\includegraphics[scale=0.5]{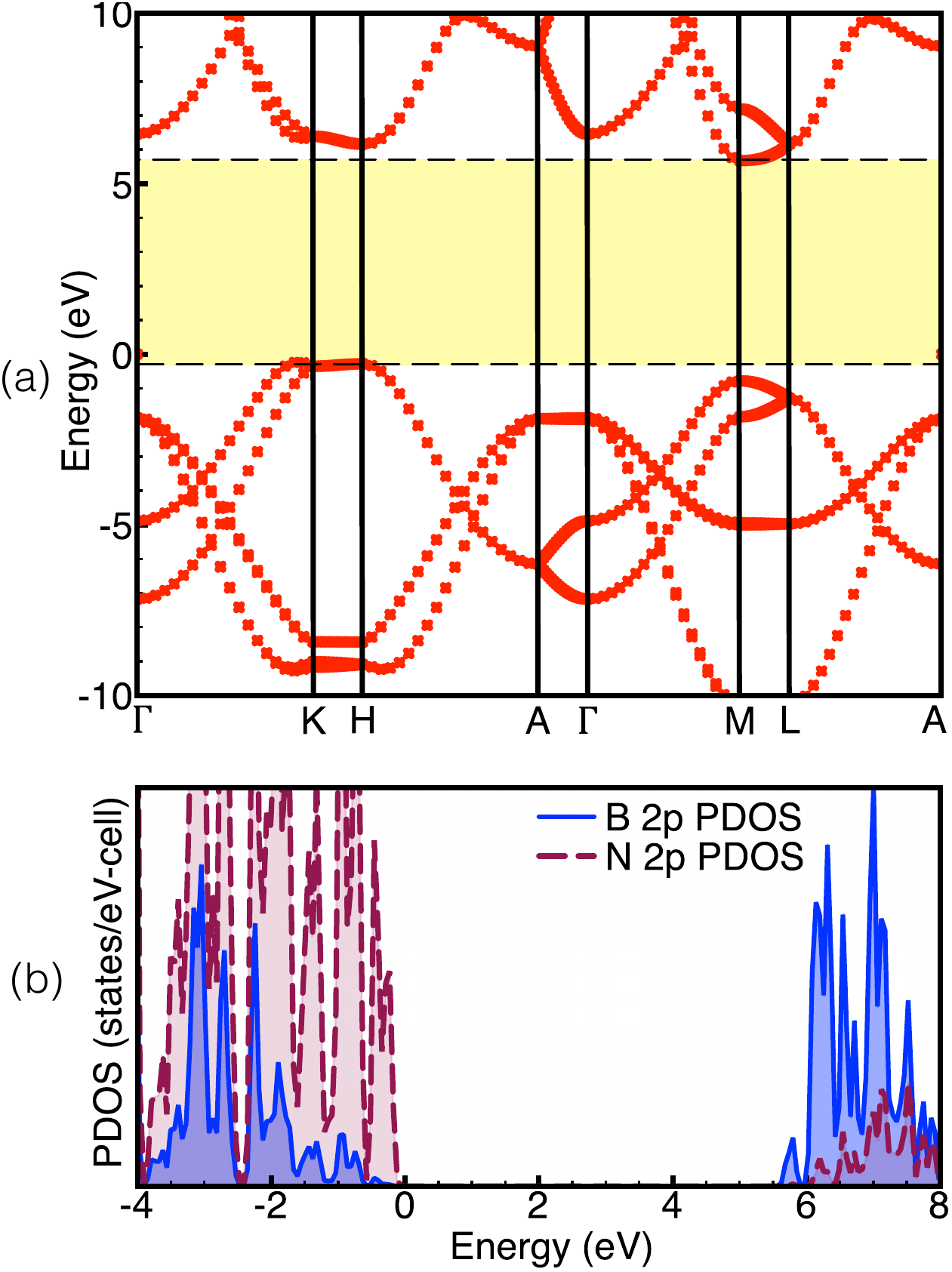}
\caption{\label{band_dos}(Color online) HSE06 calculated (a) band structure of \emph{h}-BN, (b) N-2\emph{p} and B-2\emph{p} partial DOS in \emph{h}-BN. The Fermi energy is set to zero.} 
\end{figure}

Despite the fact that there are a large number of experimental and theoretical studies on the bulk \emph{h}-BN, its electronic properties are still a matter of debate: both direct and indirect band gap with the band gap values ranging from 3.20 to 5.97 eV have been reported. For example, Watanabe \emph{et al.} showed that \emph{h}-BN has an experimental direct band gap of 5.97 eV.\cite{Watanabe2004} This result is in contradiction with the most recent quasiparticle theoretical calculations.\cite{Blase1995,Cappellini2001,Arnaud2006} Liu \emph{et al.} \cite{Liu2003} attributed such disagreement between experimental and theoretical reports to the coexistence of the sub-stable structures. Before proceeding to the results for the properties of native defects, it is worthwhile to investigate the band structure of \emph{h}-BN. Our calculations show that the VBM situates at the H point, with the conduction band minimum (CBM) locating at the M point, as shown in Fig. \ref{band_dos}(a). This leads to an indirect band gap of 5.92 eV, slightly smaller $\sim$0.4 eV than the direct band gap at the H point. Analysis of the calculated partial density of state (PDOS) presented in Fig. \ref{band_dos}(b), the upper valence band is dominated by N-2\emph{p} states, with the lower conduction band comprising mostly of B-2\emph{p} ones. In fact, as it will be shown in later, the highest occupied band (HOB) and the lowest unoccupied band (LUB) of bulk \emph{h}-BN mainly derive from N-2\emph{p}$_z$ and B-2\emph{p}$_z$ respectively [see Fig. \ref{defect_dos} (b)].

\subsection{Native defects in \emph{h}-BN}
\begin{figure}[htbp]
\centering
\includegraphics[scale=0.38]{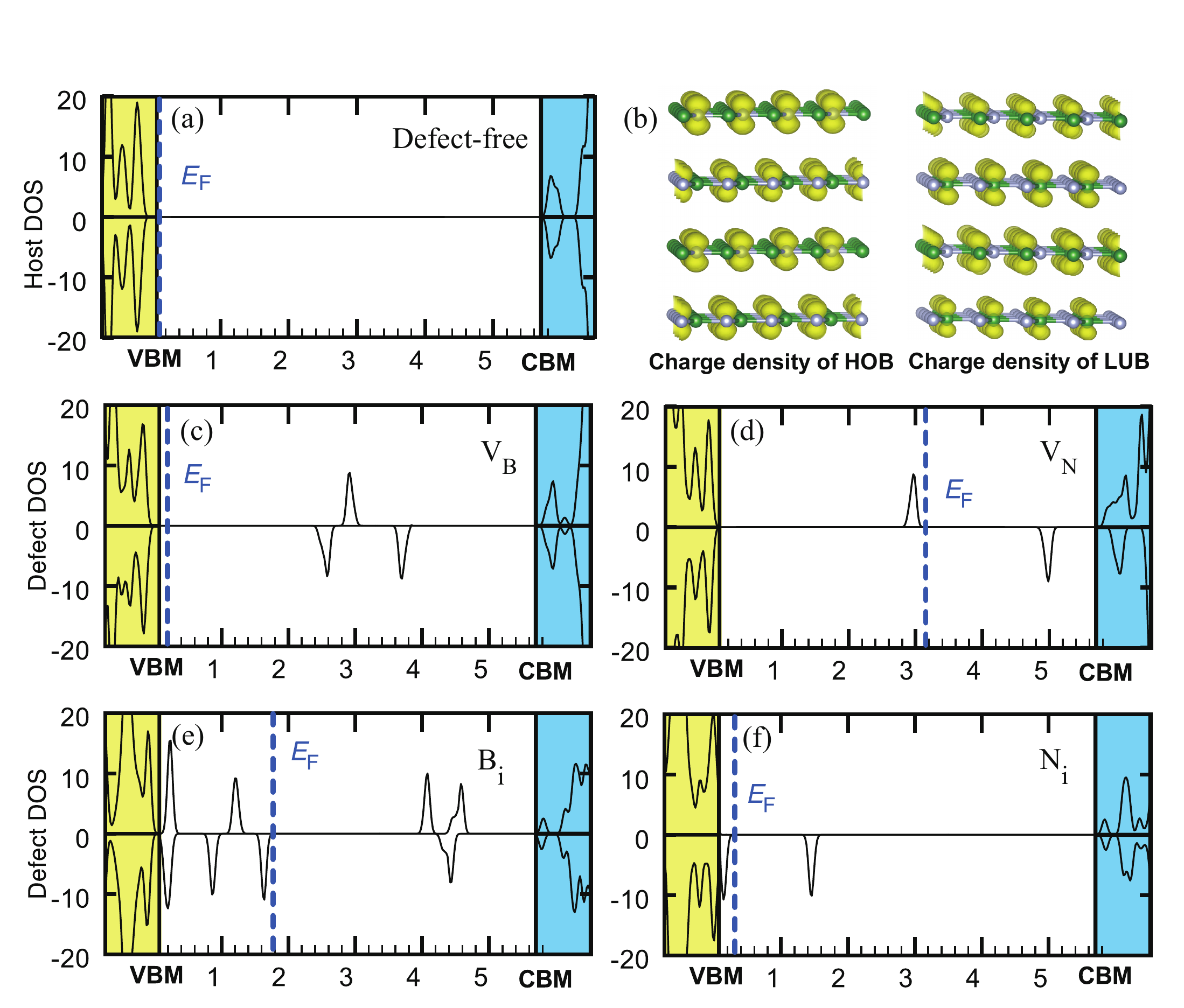}
\caption{\label{defect_dos}(Color online) (a), (\text{c}), (d), (e) and (f) HSE06-calculated defect DOS for the bulk \emph{h}-BN without native defect, with single neutral V$_\text{B}$, V$_\text{N}$, B$_\text{i}$ and N$_\text{i}$ respectively. Positive (negative) values refer to the spin-up (-down) component. The Fermi level is labeled by using a blue vertical line. (b) Band decomposed charge-density (isosurface: 0.001 \emph{e}/bohr$^3$) for the highest occupied band (HOB) and the lowest unoccupied band (LUB) for the defect-free \emph{h}-BN. Green and silver balls represent B and N atoms.} 
\end{figure}

In semiconductors and insulators, defects typically introduce levels in the band gaps of host materials. Therefore, we first examined the defect-induced DOS of the neutral B vacancy (V$_\text{B}$), N vacancy (V$_\text{N}$), interstitial B (B$_\text{i}$) and interstitial N (N$_\text{i}$), respectively. The calculated defect DOS are plotted in Fig. \ref{defect_dos}. In comparison with the DOS of defect-free system, one can note that the V$_\text{B}$ introduces one (two) unoccupied defect level(s) in the spin-up (-down) component between around 2.6 and 3.6 eV above VBM. Thus it is expected that the possible charge states of V$_\text{B}$ can vary from 0 to 3-. The V$_\text{N}$ introduces one occupied defect state at $\epsilon_{v}$+3.0 eV in the spin-up component as well as one unoccupied level at $\epsilon_{v}$+4.9 eV in the spin-down component. Its possible charge states can range from 1+ to 1-. As shown in Fig. \ref{defect_dos} (e), three occupied defects states and three unoccupied ones are introduced by the B$_\text{i}$. It should be pointed that the double-degenerate states occupied by four electrons just above CBM originate from the N-2\emph{p}$_z$ states, instead of B$_\text{i}$. The reason is attributed to the interaction between the B$_\text{i}$ and the N atoms in the adjacent BN layer. 
A wide range from 3+ to 3- can be expected for the possible charge states of B$_\text{i}$ atom. Finally, we note that one occupied defect level at $\epsilon_{v}$+0.1 eV and one unoccupied level at $\epsilon_{v}$+1.5 eV are introduced by the N$_\text{i}$. Therefore, the 1+ charge state 
is the lowest achievable charges state for the N$_\text{i}$. Adding one, and two electrons into the unoccupied levels of N$_\text{i}^{1+}$ results in the 0 and 1- charge states. Based on the results of defect induced DOS, one can find that the charge states of not all native defects can vary from 3+ to 3-.\cite{Orellana2001} Nitrogen vacancy is a typical example. 

\begin{figure}[htbp]
\centering
\includegraphics[scale=0.3]{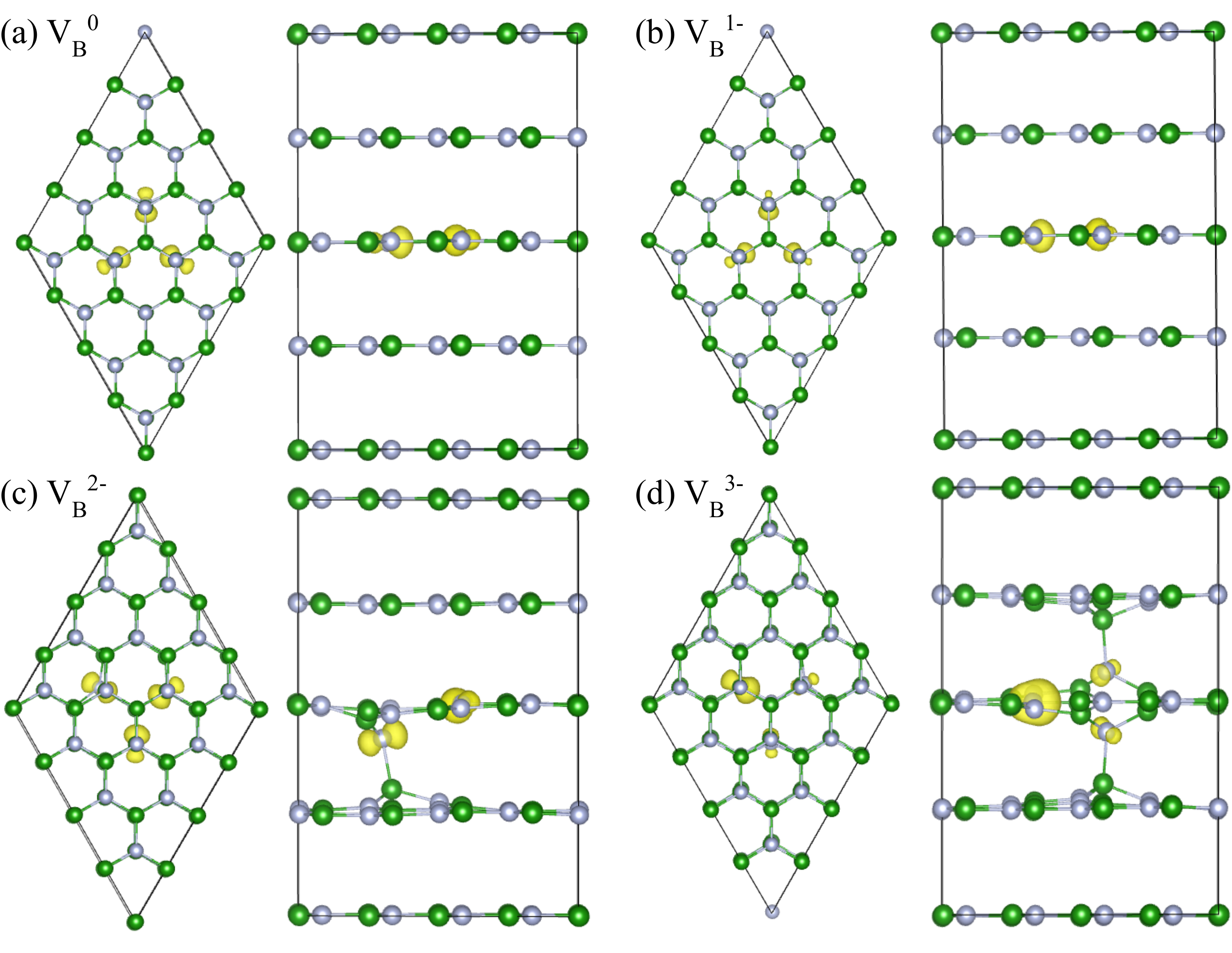}
\caption{\label{V_B}(Color online) (a), (b), (\text{c}) and (d) Top and side views of the calculated charge-density for V$_\text{B}$ in the charge state of neutral, 1-, 2- and 3-, respectively (isosurface: 0.05 \emph{e}/bohr$^3$). } 
\end{figure}

For the neutral V$_\text{B}$, it is found that its wave functions localize symmetrically over all three N neighbors and show a \emph{sp}$^2$-like hybridized character [see Fig. \ref{V_B} (a)]. No significant John-Teller distortion is observed, namely, the average N-N distances between these neighboring nitrogen atoms are almost equal to 2.7 {\AA}. The N-N distances equably reduce to 2.6 {\AA} when the V$_\text{B}$ is in the charge state of 1-. It is close to the ideal N-N distance of 2.5 {\AA} in perfect \emph{h}-BN. We attribute the decrease of N-N distance to the repulsion between B atoms and the neighboring N ones of V$_\text{B}$. It is expected that N-N distances further reduce when V$_\text{B}$ capture more extra electrons. On the other hand,  in the case of V$_\text{B}^{2-}$, the strong Coulomb repulsion between the negatively charged N atoms results in the out-of-plane distortion for one (denoted as N$_{out}$) of them. Bader charge analysis \cite{Henkelman2006} estimate that the N$_{out}$ and the other two neighboring N atoms capture about 0.95 \emph{e} and 0.44 \emph{e} per N respectively due to the existence of symmetry breaking. Consequently, the N$_{out}$ is observed to bind with the neighboring B atom in the adjacent BN layer, as shown in Fig. \ref{V_B} (\text{c}). The V$_\text{B}$ defect contributes 1 $\mu_B$ to the total magnetic moment of the system. Based on the defect levels induced by V$_\text{B}$, the total magnetic moments of the systems consisting of one V$_\text{B}^{1-}$, V$_\text{B}^{2-}$ and V$_\text{B}^{3-}$ are expected to be 2 $\mu_B$, 1 $\mu_B$ and 0 $\mu_B$. 

\begin{figure}[htbp]
\centering
\includegraphics[scale=0.3]{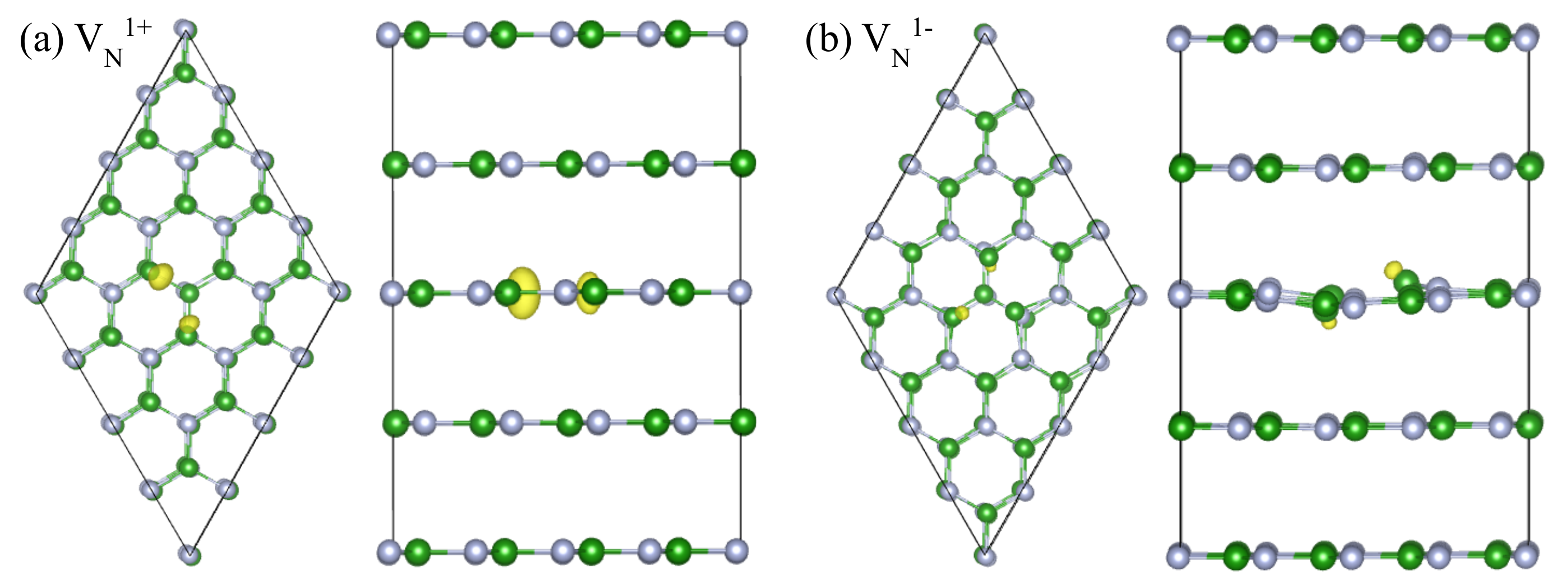}
\caption{\label{V_N}(Color online) Top and side views of the calculated charge-density for the V$_\text{N}$ defect-induced levels  (isosurface: 0.05 \emph{e}/bohr$^3$). (a) V$_\text{N}^{1+}$, (b) V$_\text{N}^{1-}$. } 
\end{figure}

In contrast to the behavior of V$_\text{B}$, the wave functions of V$_\text{N}$ defect states mainly localize on two of the three B neighbors and show a B-\emph{p}$_z$ character, as displayed in Fig.\ref{V_N}. The B-B distances between these B neighbors relax inward from 2.4 {\AA} for V$_\text{N}^{1+}$ to 2.0 {\AA} for V$_\text{N}^{1-}$.  The changes of B-B distances almost undergo within the same BN layer plane when the V$_\text{N}$ is in all possible charge states, except for the 1- state. In the latter case, a slightly out-of-plane distortion is observed for the neighboring B atoms. As displayed in Fig. \ref{defect_dos} (d), the occupied defect state at 3.0 eV above VBM leads to a magnetic moment of 1 $\mu_B$ for the neutral V$_\text{N}$. As a result, both V$_\text{N}^{1+}$ and V$_\text{N}^{1-}$ have a local magnetic moment of 0 $\mu_B$, according to the electron filling in these defect levels. 

\begin{figure}[htbp]
\centering
\includegraphics[scale=0.3]{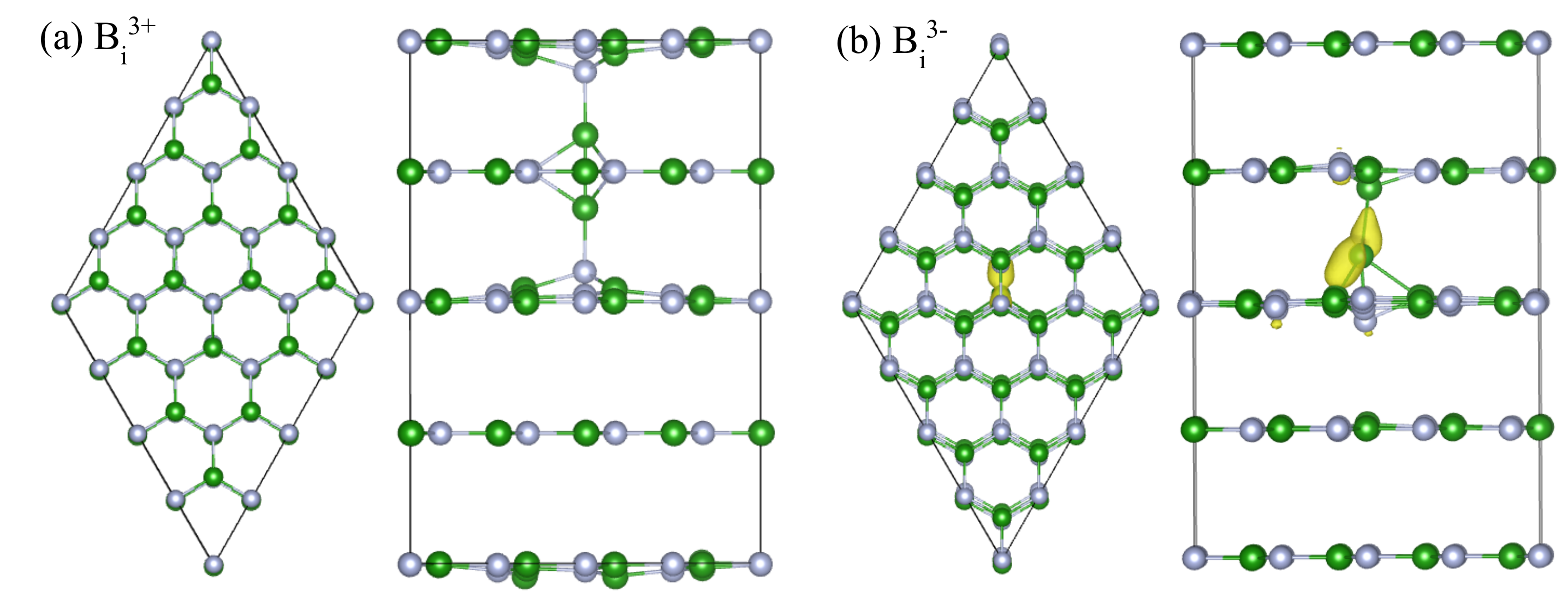}
\caption{\label{B_i}(Color online) Top and side views of the calculated charge-density (0.05 \emph{e}/bohr$^3$) for the B$_\text{i}$ defect-induced levels (isosurface:  0.05 \emph{e}/bohr$^3$). (a) B$_\text{i}^{3+}$, (b) B$_\text{i}^{3-}$. } 
\end{figure}

There are three possible interstitial sites for B/N atom occupying in bulk \emph{h}-BN. They are the hollow site above the center of a BN hexagon, the top site directly over a B atom, and the bridge site above the middle of a B-N bond. This is similar to the cases of transition-metal adsorbed on graphene layer or interstitial defects in the BN bilayer described in our previous studies.\cite{Wang201255,Wang2012}  
The optimized local structures of B$_\text{i}$ under different charge states are more complicated than the  native vacancies discussed above. As typical models, we display the relaxed local structures as well as the calculated band decomposed charge-density of B$_\text{i}^{3+}$ and B$_\text{i}^{3-}$ in Fig. \ref{B_i} (a) and (b). We note that the B$_\text{i}$ always favors the top site. It is found that a N-B-B$_i$-N vertical chain forms when the B$_\text{i}$ in the 3+ charge state. In contrast, the B$_\text{i}^{3-}$ with its three B atoms neighbors binding tetrahedrally. The wave functions of defect levels are observed to mainly localize around the B$_\text{i}$ atom. The magnetic moment of B$_\text{i}$ in any possible charge state can be deduce from the calculated value of 1 $\mu_B$ in the charge-neutral state.    

\begin{figure}[htbp]
\centering
\includegraphics[scale=0.3]{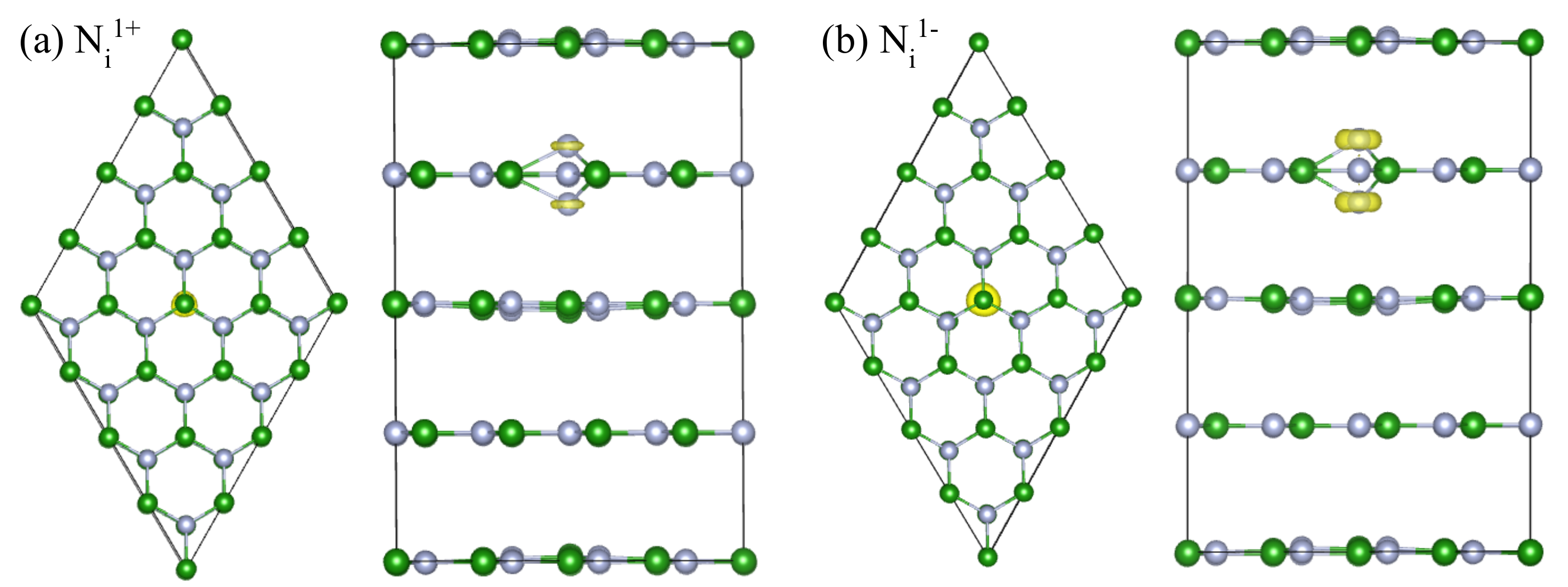}
\caption{\label{N_i}(Color online) Top and side views of the calculated charge-density for the N$_\text{i}$ defect-induced levels (isosurface:  0.05 \emph{e}/bohr$^3$). (a) N$_\text{i}^{1+}$, (b) N$_\text{i}^{1-}$.} 
\end{figure}

The N$_\text{i}$ also favors the top site. Interestingly, it is found that the N$_i$ relaxes from the ideal top site towards a lattice nitrogen atom, pushing it away from the above BN layer to form a N-N dumbbell-like structure as displayed in Fig. \ref{N_i}. One can note that the wave functions of defect levels induced by N$_\text{i}$ mainly distribute around this structure. Furthermore, the local structure of defect charge-density of N$_\text{i}$ seems to insensitive to its charge state. The calculated magnetic moment of N$_\text{i}$ in the charge states of 1+, 0, and 1- are 0 $\mu_B$, 1 $\mu_B$ and 2 $\mu_B$, respectively.

\begin{figure}[htbp]
\centering
\includegraphics[scale=0.32]{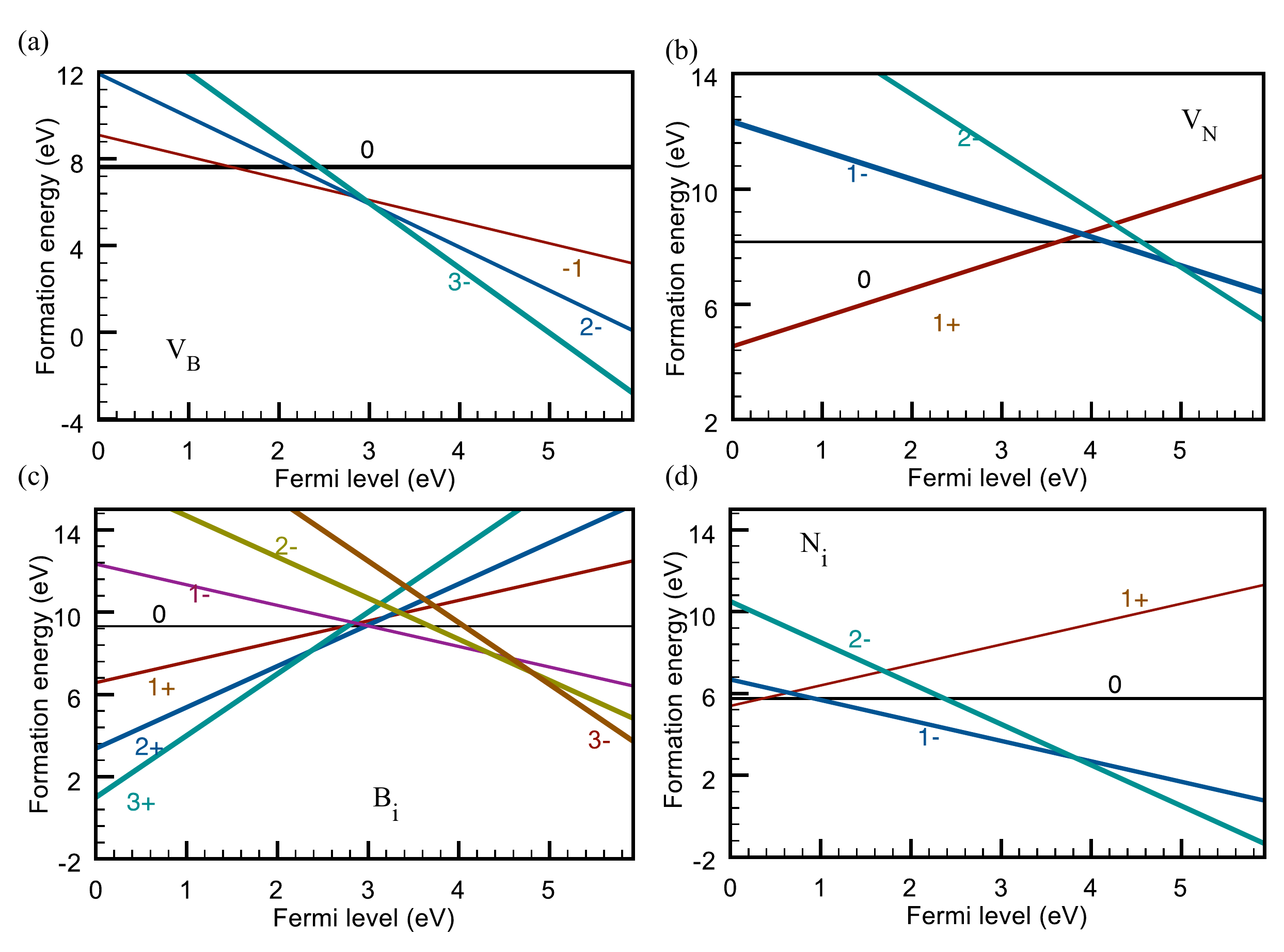}
\caption{\label{formation}(Color online) (a), (b), (\text{c}) and (d) The V$_\text{B}$, V$_\text{N}$, B$_\text{i}$ and N$_\text{i}$ formation energies as a function of the electron chemical potential under N-rich growth conditions respectively.}
\end{figure}

Since the extreme B-rich growth condition which corresponds to very low nitrogen partial pressures is probably not experimentally accessible. In our present study, we only show the formation energies of native defects under N-rich growth condition. The formation energies of V$_\text{B}$, V$_\text{N}$, B$_\text{i}$ and N$_\text{i}$ defects as a function of electron chemical potential are plotted in Fig. \ref{formation}. 
Under \emph{p}-type condition when the \emph{E}$_F$ is near VBM, the B$_\text{i}^{3+}$ with a formation energy of around -0.78 eV is found to be the most stable defect, in other words, its solubility is expected to be high under equilibrium growth conditions. The B$_\text{i}$ can act as a donor-like defect, but its calculated transition level of (3+/0) locates at 2.4 eV above VBM. This is a rather deep level, implying that it cannot compensate the \emph{p}-type conductivity in \emph{h}-BN. On the other hand, the formation energies of the other defects (V$_\text{B}$, V$_\text{N}$, and N$_\text{i}$) are high, at least 4.54 eV. This suggests that the concentration of these defect will be negligibly. When the electron chemical potential is close to CBM, the V$_\text{B}$ and  N$_\text{i}$ become more stable with respect to the V$_\text{N}$ and B$_\text{i}$. This is consistent with the experimental observation of higher concentration of nitrogen than that of boron.\cite{Shi2010} The shallowest donor levels (1-/2-) for N$_\text{i}$ and (2-/3-) for V$_\text{B}$ occurs at 2.0 eV and 3.0 eV below CBM respectively. Thus, the dominating electron killers N$_\text{i}$ and V$_\text{B}$ are also ineffective under \emph{n}-type conditions.  

\section{summary}
In summary, we have systematically studied the native defects in hexagonal BN using first-principles calculations based on hybrid density functional theory. The van der Waals interactions between BN layer were described via Grimme's DFT-D2 method. We first investigated the defect levels induced by the neutral native defects to determine the possible charge states of these defects. Then we calculated the formation energies and local structures of various native defects as a function of their possible charge states. We found that the B$_\text{i}$ will dominate in \emph{h}-BN under N-rich and \emph{p}-type conditions. On the other hand, the V$_\text{B}$ and N$_\text{i}$ become more energetically stable when the electron chemical potential is near the conduction band maximum of host. The relaxed structures of native defects were found to be strongly dependent on their charge states. Finally, based on the calculated transition energies, Our results predicted that the energetically favorable defects will not act as effective charge compensating defects under both \emph{p}- and \emph{n}-type conditions due to their ionization levels are ultra deep.

\begin{acknowledgments}
Wang acknowledges the support of the Natural National Science Foundation of Shaanxi Province (Grant No. 2013JQ1021) and the Doctoral Scientific Foundation of Xi'an University of Technology (Grant No. 108-211204). L. Ma acknowledges the support of the National Natural Science Foundation of China (Grant No. 51177133) and the Special Scientific Research Program of the Education Bureau of Shaanxi Province, China (Grant No. 2013JK1105).
\end{acknowledgments}

\nocite{*}
\bibliographystyle{aipnum4-1}

\end{document}